\documentclass[12pt]{article}
\usepackage{amsmath}
\usepackage{amsthm}
\usepackage{amssymb}
\usepackage{amscd}
\usepackage{amsfonts}
\usepackage{amsbsy}
\usepackage{graphicx}

\newtheorem{definition}{Definition}
\newtheorem{theorem}{Theorem}

\newtheorem{remark}{Remark}

\newcommand{\R}{\ensuremath{\mathbb{R}}}
\newcommand{\Z}{\ensuremath{\mathbb{Z}}}
\newcommand{\C}{\ensuremath{\mathbb{C}}}

\newcommand{\sech}{\mathop{\rm sech}\nolimits}

\title{Non-integrability of some few body problems in two degrees of freedom}
\author{Primitivo  Acosta--Hum\'anez\thanks{Departamento de Matem\'atica Aplicada I, Universidad Polit\'ecnica
de Catalunya, Spain},
Martha \'Alvarez--Ram\'\i rez and\\
 Joaqu\'{\i}n Delgado\thanks{
Departamento de Matem\'aticas,
 Universidad Aut\'onoma
Metropolitana--Iztapalapa, Mexico} }

\date{}

\begin{document}

\maketitle

\begin{abstract}
The basic theory of Differential Galois and in particular
Morales--Ramis theory is reviewed with focus in analyzing the
non--integrability of various problems of few bodies in Celestial
Mechanics. The main theoretical tools are: Morales--Ramis theorem,
the  algebrization  me\-thod of Acosta--Bl\'azquez and Kovacic's
algorithm. Morales--Ramis states that if  Hamiltonian system  has
an additional meromorphic integral in  involution in a
neighborhood of a specific solution, then  the differential Galois
group of the normal variational equations is abelian. The
algebrization method permits under general conditions to recast
the variational equation in a form suitable for its analysis by
means of Kovacic's algorithm. We apply these tools to various
examples of few body problems in Celestial Mechanics: (a) the
elliptic restricted three body  in the plane with collision of the
primaries; (b) a general Hamiltonian system of two degrees of
freedom with homogeneous potential of degree $-1$; here we perform
McGehee's blow up and obtain the normal variational equation in
the form of an hypergeometric equation. We recover Yoshida's
criterion for non--integrability. Then we contrast two methods to
compute the Galois group: the well known, based in the
Schwartz--Kimura table, and the lesser based in Kovacic's
algorithm. We apply these methodology to three problems: the
rectangular four body problem, the anisotropic Kepler problem and
two uncoupled Kepler problems in the line; the last two depend on
a mass parameter, but while in the anisotropic problem it is
integrable for only two values of the parameter, the two uncoupled
Kepler problems is completely integrable for all values of the
masses.\\

\noindent\textbf{Keywords and Phrases.} Algebrization method,
Differential Galois Theory, Celestial mechanics, Kovacic's
algorithm, Kimura's theorem, Morales-Ramis theory, $n-$body
problems, non-integrability.\\

\noindent\textbf{AMS Subject Classification.} 37J30; 12H05; 34M15;
70H07; 70F10; 47J30

\end{abstract}

\section{Introduction}

In this paper we analyze the integrability of some Hamiltonian
systems of two degrees of freedom related with few body problems.
This can be made through the analysis of the linearization of the
Hamiltonian system, that is, variational equations and normal
variational equations. In 1982 Ziglin (\cite{ziglin}) proved a
non-integrability theorem using the constraints imposed on the
monodromy group of the normal variational equations along some
integral curve by the existence of some first integrals. This is a
result about branching of solutions: the monodromy group express
the ramification of the solutions of the normal variational
equation in the complex domain.

We consider a {\it complex} analytic symplectic manifold $M$ of
dimension $2n$ and a holomorphic hamiltonian system $X_H$ defined
over it. Let $\Gamma $ be the Riemann surface corresponding to an
integral curve $z=z(t)$ (which is not an equilibrium point)  of
the vector field $X_H$. Then we can write the variational
equations (VE) along $\Gamma $,

\[
\dot\eta=\frac{\partial X_H}{\partial x}(z(t))\eta.
\]
Using the linear first integral $dH(z(t))$ of the VE it is
possible to reduce this variational equation (i.e. to rule out one
degree of freedom) and to obtain the so called normal variational
equation (NVE) that, in some adequate coordinates, we can write,

\[
\dot\xi=JS(t)\xi,
\]
where, as usual,
\[ J=\left(\begin{array}{cc} 0 & I\\
-I & 0 \end{array} \right) \] is the square matrix of the
symplectic form. (Its dimension is $2(n- 1)$).

In general if, including the hamiltonian, there are $k$ analytical
first integrals independent over $\Gamma$ and in involution, then,
in a similar way, we can reduce the number of degrees of freedom
of the VE by $k$. The resulting equation, which admits $n-k$
degrees of freedom, is also called the normal variational equation
(NVE). Then we have the following result (\cite{ziglin}).
\\

\noindent\textbf{Theorem (Ziglin).}\emph{ Suppose that the
hamiltonian system admits $n-k$ additional analytical first
integrals, independent over a neighborhood of $\Gamma$ (but not
necessarily on $\Gamma$ itself) We assume moreover that the
monodromy group of the NVE contain a non-resonant transformation
$g$. Then, any other element of the monodromy group of the NVE
send eigendirections of $g$ into eigendirections of $g$.}
\\

We recall that a linear transformation $g\in Sp(m,\C)$ (the
monodromy group is contained in the symplectic group) is resonant
if there exists integers $r_1,...,r_m$ such that
$\lambda_1^{r_1}\cdots\lambda^{r_m}=1$ (where we denoted by
$\lambda_i$ the eigenvalues of $g$).

Later, Morales and Ramis in 2001 improved the Ziglin's result by
means of differential Galois theory (see \cite{moralesramis} and
see also \cite{Morales}), arising in this way the so-called
\emph{Morales-Ramis Theory}. This theory will be explained in
section \ref{morrami} of this paper.

There are a lot of papers and books devoted to analyze three body
problems (see \cite{simon} and references therein). Therefore, an
special kind of three body problem is the so-called \emph{Sitnikov
problem}, which has been deeply analyzed using Morales-Ramis
theory in \cite{ac,Morales,moralesramisII}. Another cases of three
body problems has been studied, also by means of Morales-Ramis
theory, in \cite{Boucher,BoucherWeil}. There are a lot cases in
which the variational equation falls in Riemann differential
equation or hypergeometric differential equation. In this cases
has been used satisfactory the Kimura-Schwartz table (see
\cite{kimura}), which was improved by Morales in \cite{Morales}.

In this paper, we analyze the non-integrability of some celestial
mechanics problems such as the collinear restricted elliptic
three-body problem, rectangular 4 body problem and the anisotropic
Kepler problem. The approach used here is by means of Morales
Ramis theory contrasting the Kovacic's algorithm with Kimura's
theorem, but obtaining the same results.

\section{Differential Galois Theory}\label{galois}

Our theoretical framework consists of a well-established crossroads of
Dynamical Systems theory, Algebraic Geometry and Differential
Algebra. See \cite{Morales} or \cite{SingerVanderput} for further
information and details. Given a linear differential system with
coefficients in $\mathbb{C}\!\left(t\right)$,
\begin{equation} \label{HODE}
\dot{{z}}=A\left( t\right) {z},
\end{equation}
a differential field $L\supset \mathbb{C}\!\left(t\right)$ exists,
unique up to $\mathbb{C}\!\left(t\right)$-isomorphism, which
contains all entries of a fundamental matrix
$\Psi=\left[{\psi}_1,\dots,{\psi}_n\right]$ of \eqref{HODE}.
Moreover, the group of differential automorphisms of this field
extension, called the \emph{differential Galois group} of
\eqref{HODE}, is an algebraic group $G$ acting over the
$\mathbb{C}$-vector space $\left\langle {\psi}_1,\dots,{\psi}_n
\right\rangle$ of solutions of \eqref{HODE} and containing the
monodromy group of \eqref{HODE}.

It is worth recalling that the integrability of a linear system
\eqref{HODE} is equivalent to the solvability of the identity
component $G^0$ of the differential Galois group $G$ of
\eqref{HODE} -- in other words, equivalent to the \emph{virtual
solvability} of $G$.

It is well established (e.g. \cite{ac,mo1}) that any linear
differential equation system with coefficients in a differential
field $K$
\begin{equation}\label{sisor1} \frac d{dt}\left(\begin{array}{c} \xi_1 \\
 \xi_2
\end{array}\right) =\left(\begin{array}{cc} a(t) & b(t)  \\ c(t) & d(t)
  \end{array}\right)\left(\begin{array}{c} \xi_1 \\
 \xi_2
\end{array}\right),
\end{equation}
by means of an elimination process, is equivalent to the
second-order equation \small{\begin{equation}\label{sisor2} \ddot
\xi-\left(a(t)+d(t)+\frac {\dot b(t)} {b(t)}\right)\dot \xi
-\left(\dot a(t)+b(t)c(t)-a(t)d(t)-\frac{a(t)\dot
b(t)}{b(t)}\right)\xi=0,
\end{equation}}
where $\xi:=\xi_1$. Furthermore, any equation of the form $\ddot
z-2p\dot z-qz=0$, can be transformed, through the change of
variables $z=ye^{{\int}p}$, into $\ddot y=-ry$, $r$ satisfying the
Riccati equation $\dot p=r+q+p^2$. This change is useful since it
restricts the study of the Galois group of $\ddot y=-ry$ to that
of the algebraic subgroups of $\textrm{SL}(2,\mathbb{C})$. This
last procedure will be used later in section \emph{algorithmic
approach}.

A natural question which now arises is to determine what happens
if the coefficients of the differential equation are not all
rational. A new method was developed in \cite{acbl}, in order to
transform a linear differential equation of the form $\ddot
x=r(t)x$, with transcendental or algebraic non-rational
coefficients, into its algebraic form -- that is, into a
differential equation with rational coefficients. This is called
the \emph{algebrization method} and is based on the concept of
\emph{Hamiltonian change of variables} \cite{acbl}. Such a change
is derived from the solution of a one-degree-of-freedom classical
Hamiltonian.
\\

\begin{definition}[Hamiltonian change of variables]\label{defi2} A change of variables $\tau=\tau(t)$ is called \textbf{Hamiltonian} if
$(\tau(t), \dot\tau(t))$ is a solution curve of the autonomous
Hamiltonian system $X_H$ with Hamiltonian function
$$H=H(\tau,p)={\frac{p^2}{2}}+\widehat{V}(\tau),\textit{ for some } \widehat{V}\in \mathbb
C(\tau).$$
\end{definition}

\begin{theorem}[Acosta-Bl\'azquez algebrization method \cite{acbl}]\label{pr2}
Equation $\ddot x=r(t)x$ is algebrizable by means of a Hamiltonian change of variables
$\tau=\tau(t)$ if, and only if, there exist $f,\alpha$ such that
${\frac{d}{d\tau}}\left(\ln \alpha\right),{\frac{f}{\alpha}}\in \mathbb{C}(\tau),$
where $$f(\tau(t))=r(t),\quad \alpha(\tau)=2(H-\widehat{V}(\tau))=(\dot\tau)^2.$$
Furthermore, the algebraic form of
$\ddot x=r(t)x$ is
\begin{equation}\label{equ4}
\frac{d^2 x }{d\tau^2}+\left(\frac{1}{2}\frac{d}{d\tau}\ln \alpha\right)\frac{dx}{d\tau}-\left(\frac{f}{\alpha}\right)x=0.  \quad\square
\end{equation}
\end{theorem}

The next intended step, once a differential equation has been
algebrized, is studying its Galois group and, as a causal
consequence, its integrability. Concerning the latter, and in
virtue of the invariance of the identity component of the Galois
group by finite branched coverings of the independent variable
(Morales-Ruiz and Ramis, \cite[Theorem 5]{moralesramis}), it was
proven in \cite[Proposition 1]{acbl} that the identity component
of \emph{the Galois group is preserved in the algebrization
mechanism}.

The final step is analyzing the behavior of $t=\infty$ (or $\tau=\infty$) by studying the behavior of $\eta=0$
through the change of variables $\eta=1/t$ (or $\eta=1/\tau$) in the transformed differential equation, i.e.
$t=\infty$ (or $\tau=\infty$) is an ordinary point (resp. a regular singular point, an irregular singular point) of
the original differential equation if, and only if, $\eta=0$ is one such point for the transformed differential equation.

\section{Morales-Ramis Theory}\label{morrami}
\emph{Everything is considered in the complex analytical setting from now on.} The heuristics of
the titular theory rest on the following general principle: if we assume system
\begin{equation}\label{DS} {\dot{z}}=X\left({z}\right)
\end{equation}
``integrable" in some reasonable sense, then the corresponding
variational equations along any integral curve
$\Gamma=\left\{\widehat{{z}}\left(t\right):t\in I\right\}$ of
\eqref{DS}, defined in the usual manner
\begin{equation}\label{vegamma}\tag{$\mathrm{VE}_{\Gamma}$}
\dot{{\xi}}=X'\left(\widehat{{z}}\left(t\right)\right){\xi},
\end{equation}
must be also integrable -- in the Galoisian sense of the last
paragraph in \ref{galois}. We assume $\Gamma$, a Riemann surface,
may be locally parameterized in a disc $I$ of the complex plane;
we may now complete $\Gamma$ to a new Riemann surface
$\overline{\Gamma}$, as detailed in \cite[\S 2.1]{moralesramis}
(see also \cite[\S 2.3]{Morales}), by adding equilibrium points,
singularities of the vector field and possible points at infinity.
Linearization  defines a linear connection over
$\overline{\Gamma}$ called the variational connection
$\mathrm{VE}_{\overline{\Gamma}}$ and
$\mathrm{Gal}\left(\mathrm{VE}_{\overline{\Gamma}}\right)$ is its
Galois differential group which contains  the Zariski closure of
the monodromy group
$\mathrm{Mon}\left(\mathrm{VE}_{\overline{\Gamma}}\right)$. In
practice the normal variational equations
$NVE_{\overline{\Gamma}}=TM|_{\Gamma}/T\Gamma$ are analyzed, the
variational equation along the solution being reducible.

The aforementioned ``reasonable'' sense in which to define
integrability if system \eqref{DS} is \emph{Hamiltonian} is
obviously the one given by the Liouville-Arnold Theorem (see
\cite{AbrahamMarsden,Arnold,Whittaker}), and thus the above
general principle does have an implementation:
\\

\begin{theorem}[J. Morales-Ruiz \& J.-P. Ramis, 2001] \label{moralesramis}
Let $H$ be an $n$-degree-of-freedom Hamiltonian having $n$
independent rational or meromorphic first integrals in pairwise
involution, defined on a neighborhood of an integral curve
$\overline{\Gamma}$. Then, the identity component
$\mathrm{Gal}\left(\mathrm{VE}_{\overline{\Gamma}}\right)^0$ is an
abelian group (i.e.
$\mathrm{Gal}\left(\mathrm{VE}_{\overline{\Gamma}}\right)$ is
\emph{virtually abelian}).
\end{theorem}

The disjunctive between \emph{meromorphic} and \emph{rational}
Hamiltonian integrability in Theorem \ref{moralesramis} is related
to the status of $t=\infty$ as a singularity for the normal
variational equations. More specifically, and besides the
non-abelian character of the identity component of the Galois
group, in order to obtain Galoisian obstructions to the
\emph{meromorphic} integrability of $H$ the point at infinity must
be a regular singular point of \eqref{vegamma} (for example
Hypergeometric and Riemann differential equations). On the other
hand, for there to be an obstruction to complete sets of
\emph{rational} first integrals, $t=\infty$ must be a irregular
singular point. See \cite[Corollary 8]{moralesramis} or
\cite[Theorem 4.1]{Morales} for a precise statement and a proof.
\\

Different notions of integrability correspond to  classes of
admissible first integrals, for instance rational, meromorphic, algebraic,
smooth, etc. For non--integrability within the real--analytic
realm, a popular testing is Melnikov integral whose isolated zeros
give transversal homoclinic intersections and in under proper
hypothesis, chaos. Galoisian obstruction to integrability based in
Morales-Ramis theory has shown to be equivalent to the presence of
isolated zeros of Melnikov integral, for a class of Hamiltonian
systems with two degrees of freedom with saddle centers (see
\cite{MoralesSplitting,yagasaki}). Also high order variational
equations have been studied in this context (see
\cite{MorRamSimo}).
\\

\begin{remark}
 In order to analyze normal variational equations, a standard
procedure is using \textsc{Maple}, and especially commands
\texttt{dsolve} and \texttt{kovacicsols}. Whenever the command
\texttt{kovacicsols} yields an output ``\texttt{[ ]}", it means
that the second-order linear differential equation being
considered has no Liouvillian solutions, and thus its Galois group
is virtually non-solvable. For equations of the form $\ddot y=ry$
with $r\in\mathbb{C}(x)$ the only virtually non-solvable group is
$\mathrm{SL}(2,\mathbb{C})$. In some cases, moreover,
\texttt{dsolve} makes it possible to obtain the solutions in terms
of special functions such as \emph{Airy functions}, \emph{Bessel
functions} and \textbf{hypergeometric functions}, among others
(\cite{Abramowitz}). There is a number of second-order linear
equations whose coefficients are not rational, and whose solutions
\textsc{Maple} cannot find by means of the commands
\texttt{dsolve} and \texttt{kovacicsols} alone; this problem, in
some cases, can be solved by the stated algebrization procedure.
Another difficulty is when appears parameters in the differential
equation, then almost always \texttt{kovacicsols} wrong, for this
reason we present in following section the Kovacic's algorithm to be used
later.
\end{remark}

\section{Kovacic's Algorithm}
This algorithm is devoted to solve the RLDE (reduced linear
differential equation) $\xi''=r\xi$ and is based on the algebraic
subgroups of $\mathrm{SL}(2,\mathbb{C}).$ For more details see
\cite{Kov}. Improvements for this algorithm are given in
\cite{UlmerWeil}, where it is not necessary to reduce the
equation. Another improvement is given in \cite{dulo}, which is a
compact version to implement in computer systems. Here, we follow
the original version given by Kovacic in \cite{Kov}, which is the
same version given in \cite{acbl}.

\begin{theorem}\label{subgroups} Let $G$ be an algebraic subgroup of $\mathrm{SL}(2,\mathbb{C})$.  Then
one of the following four cases can occur.
\begin{enumerate}
\item $G$ is triangularizable.
\item $G$ is conjugate to a subgroup of infinite dihedral group (also called meta-abelian group)
and case 1 does not hold.
\item Up to conjugation $G$ is one of the following finite groups: Tetrahedral group, Octahedral
group or Icosahedral group, and cases 1 and 2 do not hold.
\item $G = \mathrm{SL}(2,\mathbb{C})$.
\end{enumerate}
\end{theorem}

Each case in Kovacic's algorithm is related with each one of the
algebraic subgroups of $\mathrm{SL}(2,\mathbb{C})$ and the
associated Riccatti equation
$$\theta^{\prime}=r-\theta ^{2}=\left( \sqrt{r}-\theta\right)
\left(  \sqrt{r}+\theta\right),\quad\theta={\xi'\over \xi}.$$

According to Theorem \ref{subgroups}, there are four cases in
Kovacic's algorithm. Only for cases 1, 2 and 3 we can solve the
differential equation the RLDE, but for the case 4 we have not
Liouvillian solutions for the RLDE. It is possible that Kovacic's
algorithm can provide us only one solution ($\xi_1$), so that we
can obtain the second solution ($\xi_2$) through
\begin{equation}\label{second}
\xi_2=\xi_1\int\frac{dx}{\xi_1^2}.
\end{equation}

{\bf\large Notations.} For the RLDE given by
$${d^2\xi\over dx^2}=r\xi,\qquad r={s\over t},\quad s,t\in \mathbb{C}[x],$$
we use the following notations.
\begin{enumerate}
\item Denote by $\Upsilon'$ be
the
set of (finite) poles of $r$, $\Upsilon^{\prime}=\left\{  c\in\mathbb{C}%
:t(c)=0\right\}$.

\item Denote by
$\Upsilon=\Upsilon^{\prime}\cup\{\infty\}$.
\item By the order of $r$ at
$c\in \Upsilon'$, $\circ(r_c)$, we mean the multiplicity of $c$ as
a pole of $r$.

\item By the order of $r$ at $\infty$, $\circ\left(
r_{\infty}\right) ,$ we mean the order of $\infty$ as a zero of
$r$. That is $\circ\left( r_{\infty }\right) =deg(t)-deg(s)$.

\end{enumerate}
\subsection{The four cases\label{four-cases}}

{\bf\large Case 1.} In this case $\left[ \sqrt{r}\right] _{c}$ and
$\left[ \sqrt{r}\right] _{\infty}$ means the Laurent series of
$\sqrt{r}$ at $c$ and the Laurent series of $\sqrt{r}$ at $\infty$
respectively. Furthermore, we define $\varepsilon(p)$ as follows:
if $p\in\Upsilon,$ then $\varepsilon\left( p\right) \in\{+,-\}.$
Finally, the complex numbers $\alpha_{c}^{+},\alpha_{c}^{-},\alpha_{\infty}%
^{+},\alpha_{\infty}^{-}$ will be defined in the first step. If
the differential equation has not poles it only can fall in this
case.
\medskip

{\bf Step 1.} Search for each $c \in \Upsilon'$ and for $\infty$
the corresponding situation as follows:

\medskip

\begin{description}

\item[$(c_{0})$] If $\circ\left(  r_{c}\right)  =0$, then
$$\left[ \sqrt {r}\right] _{c}=0,\quad\alpha_{c}^{\pm}=0.$$

\item[$(c_{1})$] If $\circ\left(  r_{c}\right)  =1$, then
$$\left[ \sqrt {r}\right] _{c}=0,\quad\alpha_{c}^{\pm}=1.$$

\item[$(c_{2})$] If $\circ\left(  r_{c}\right)  =2,$ and $$r= \cdots
+ b(x-c)^{-2}+\cdots,\quad \text{then}$$
$$\left[ \sqrt {r}\right]_{c}=0,\quad \alpha_{c}^{\pm}=\frac{1\pm\sqrt{1+4b}}{2}.$$

\item[$(c_{3})$] If $\circ\left(  r_{c}\right)  =2v\geq4$, and $$r=
(a\left( x-c\right)  ^{-v}+...+d\left( x-c\right)
^{-2})^{2}+b(x-c)^{-(v+1)}+\cdots,\quad \text{then}$$ $$\left[
\sqrt {r}\right] _{c}=a\left( x-c\right) ^{-v}+...+d\left(
x-c\right) ^{-2},\quad\alpha_{c}^{\pm}=\frac{1}{2}\left(
\pm\frac{b}{a}+v\right).$$

\item[$(\infty_{1})$] If $\circ\left(  r_{\infty}\right)  >2$, then
$$\left[\sqrt{r}\right]  _{\infty}=0,\quad\alpha_{\infty}^{+}=0,\quad\alpha_{\infty}^{-}=1.$$

\item[$(\infty_{2})$] If $\circ\left(  r_{\infty}\right)  =2,$ and
$r= \cdots + bx^{2}+\cdots$, then $$\left[
\sqrt{r}\right]  _{\infty}=0,\quad\alpha_{\infty}^{\pm}=\frac{1\pm\sqrt{1+4b}%
}{2}.$$

\item[$(\infty_{3})$] If $\circ\left(  r_{\infty}\right) =-2v\leq0$,
and
$$r=\left( ax^{v}+...+d\right)  ^{2}+ bx^{v-1}+\cdots,\quad \text{then}$$
$$\left[  \sqrt{r}\right]  _{\infty}=ax^{v}+...+d,\quad
\text{and}\quad \alpha_{\infty}^{\pm }=\frac{1}{2}\left(
\pm\frac{b}{a}-v\right).$$
\end{description}
\medskip

{\bf Step 2.} Find $D\neq\emptyset$ defined by
$$D=\left\{
m\in\mathbb{Z}_{+}:m=\alpha_{\infty}^{\varepsilon
(\infty)}-%
{\displaystyle\sum\limits_{c\in\Upsilon^{\prime}}}
\alpha_{c}^{\varepsilon(c)},\forall\left(  \varepsilon\left(
p\right) \right)  _{p\in\Upsilon}\right\}  .$$ If $D=\emptyset$,
then we should start with the case 2. Now, if $\#D>0$, then for
each $m\in D$ we search $\omega$ $\in\mathbb{C}(x)$ such that
$$\omega=\varepsilon\left(
\infty\right)  \left[  \sqrt{r}\right]  _{\infty}+%
{\displaystyle\sum\limits_{c\in\Upsilon^{\prime}}}
\left(  \varepsilon\left(  c\right)  \left[  \sqrt{r}\right]  _{c}%
+{\alpha_{c}^{\varepsilon(c)}}{(x-c)^{-1}}\right).$$
\medskip

{\bf Step 3}. For each $m\in D$, search for a monic polynomial
$P_m$ of degree $m$ with
$$P_m'' + 2\omega P_m' + (\omega' + \omega^2 - r) P_m = 0.$$

If success is achieved then $\xi_1=P_m e^{\int\omega}$ is a
solution of the differential equation the RLDE.  Else, Case 1
cannot hold.
\bigskip

{\bf\large Case 2.}  Search for each $c \in \Upsilon'$ and for
$\infty$ the corresponding situation as follows:
\medskip

{\bf Step 1.} Search for each $c\in\Upsilon^{\prime}$ and $\infty$
the sets $E_{c}\neq\emptyset$ and $E_{\infty}\neq\emptyset.$ For
each $c\in\Upsilon^{\prime}$ and for $\infty$ we define
$E_{c}\subset\mathbb{Z}$ and $E_{\infty}\subset\mathbb{Z}$ as
follows:
\medskip

\begin{description}
\item[($c_1$)] If $\circ\left(  r_{c}\right)=1$, then $E_{c}=\{4\}$

\item[($c_2$)] If $\circ\left(  r_{c}\right)  =2,$ and $r= \cdots +
b(x-c)^{-2}+\cdots ,\ $ then $$E_{c}=\left\{
2+k\sqrt{1+4b}:k=0,\pm2\right\}.$$

\item[($c_3$)] If $\circ\left(  r_{c}\right)  =v>2$, then $E_{c}=\{v\}$

\item[$(\infty_{1})$] If $\circ\left(  r_{\infty}\right)  >2$, then
$E_{\infty }=\{0,2,4\}$

\item[$(\infty_{2})$] If $\circ\left(  r_{\infty}\right)  =2,$ and
$r= \cdots + bx^{2}+\cdots$, then $$E_{\infty }=\left\{
2+k\sqrt{1+4b}:k=0,\pm2\right\}.$$

\item[$(\infty_{3})$] If $\circ\left(  r_{\infty}\right)  =v<2$,
then $E_{\infty }=\{v\}$
\medskip
\end{description}

{\bf Step 2.} Find $D\neq\emptyset$ defined by
$$D=\left\{
m\in\mathbb{Z}_{+}:\quad m=\frac{1}{2}\left(  e_{\infty}-
{\displaystyle\sum\limits_{c\in\Upsilon^{\prime}}} e_{c}\right)
,\forall e_{p}\in E_{p},\text{ }p\in\Upsilon\right\}.$$ If
$D=\emptyset,$ then we should start the case 3. Now, if $\#D>0,$
then for each $m\in D$ we search a rational function $\theta$
defined by
$$\theta=\frac{1}{2}
{\displaystyle\sum\limits_{c\in\Upsilon^{\prime}}}
\frac{e_{c}}{x-c}.$$
\medskip

{\bf Step 3.} For each $m\in D,$ search a monic polynomial $P_m$
of degree $m$, such that
$$P_m^{\prime\prime\prime}+3\theta
P_m^{\prime\prime}+(3\theta^{\prime}+3\theta
^{2}-4r)P_m^{\prime}+\left(  \theta^{\prime\prime}+3\theta\theta^{\prime}%
+\theta^{3}-4r\theta-2r^{\prime}\right)P_m=0.$$ If $P_m$ does not
exist, then Case 2 cannot hold. If such a polynomial is found, set
$\phi = \theta + P'/P$ and let $\omega$ be a solution of
$$\omega^2 + \phi \omega + {1\over2}\left(\phi' + \phi^2 -2r\right)=
0.$$

Then $\xi_1 = e^{\int\omega}$ is a solution of the differential
equation the RLDE.
\bigskip

{\bf\large Case 3.} Search for each $c \in \Upsilon'$ and for
$\infty$ the corresponding situation as follows:
\medskip

{\bf Step 1.} Search for each $c\in\Upsilon^{\prime}$ and $\infty$
the sets $E_{c}\neq\emptyset$ and $E_{\infty}\neq\emptyset.$ For
each $c\in\Upsilon^{\prime}$ and for $\infty$ we define
$E_{c}\subset\mathbb{Z}$ and $E_{\infty}\subset\mathbb{Z}$ as
follows:
\medskip

\begin{description}

\item[$(c_{1})$] If $\circ\left(  r_{c}\right)  =1$, then
$E_{c}=\{12\}$

\item[$(c_{2})$] If $\circ\left(  r_{c}\right)  =2,$ and $r= \cdots +
b(x-c)^{-2}+\cdots$, then
\begin{displaymath}
E_{c}=\left\{ 6+k\sqrt{1+4b}:\quad
k=0,\pm1,\pm2,\pm3,\pm4,\pm5,\pm6\right\}.
\end{displaymath}

\item[$(\infty)$] If $\circ\left(  r_{\infty}\right)  =v\geq2,$ and $r=
\cdots + bx^{2}+\cdots$, then
$$E_{\infty }=\left\{
6+{12k\over n}\sqrt{1+4b}:\quad
k=0,\pm1,\pm2,\pm3,\pm4,\pm5,\pm6\right\},\quad n\in\{4,6,12\}.$$
\medskip
\end{description}

{\bf Step 2.} Find $D\neq\emptyset$ defined by
$$D=\left\{
m\in\mathbb{Z}_{+}:\quad m=\frac{n}{12}\left(
e_{\infty}-{\displaystyle\sum\limits_{c\in\Upsilon^{\prime}}}
e_{c}\right)  ,\forall e_{p}\in E_{p},\text{
}p\in\Upsilon\right\}.$$ In this case we start with $n=4$ to
obtain the solution, afterwards $n=6$ and finally $n=12$. If
$D=\emptyset$, then the differential equation has not Liouvillian
solution because it falls in the case 4. Now, if $\#D>0,$ then for
each $m\in D$ with its respective $n$, search a rational function
$$\theta={n\over 12}{\displaystyle\sum\limits_{c\in\Upsilon^{\prime}}}
\frac{e_{c}}{x-c}$$ and a polynomial $S$ defined as $$S=
{\displaystyle\prod\limits_{c\in\Upsilon^{\prime}}} (x-c).$$

{\bf Step 3}. Search for each $m\in D$, with its respective $n$, a
monic polynomial $P_m=P$ of degree $m,$ such that its coefficients
can be determined recursively by
$$\bigskip P_{-1}=0,\quad P_{n}=-P,$$
$$P_{i-1}=-SP_{i}^{\prime}-\left( \left( n-i\right)
S^{\prime}-S\theta\right)  P_{i}-\left( n-i\right)  \left(
i+1\right)  S^{2}rP_{i+1},$$ where $i\in\{0,1\ldots,n-1,n\}.$ If
$P$ does not exist, then the differential equation has not
Liouvillian solution because it falls in Case 4. Now, if $P$
exists search $\omega$ such that $$
{\displaystyle\sum\limits_{i=0}^{n}} \frac{S^{i}P}{\left(
n-i\right)  !}\omega^{i}=0,$$ then a solution of the differential
equation the RLDE is given by $$\xi=e^{\int \omega},$$ where
$\omega$ is solution of the previous polynomial of degree $n$.
\bigskip

\subsection{Some remarks on Kovacic's algorithm}
Along this section we assume that the RLDE falls only in one of
the four cases.
\begin{remark}[Case 1]\label{rkov2}
If the RLDE falls in case 1, then its Galois group is given by one
of the following groups:

\begin{description}

\item[I1] $e$ when the algorithm provides two
rational solutions or only one rational solution and the second
solution obtained by \eqref{second} has not logarithmic term.
$$e=\left\{\begin{pmatrix}1&0\\0&1\end{pmatrix}\right\},$$
this group is connected and abelian.

\item[I2] $\mathbb{G}_k$ when the algorithm provides only one
algebraic solution $\xi$ such that $\xi^k\in\mathbb{C}(x)$ and
$\xi^{k-1}\notin\mathbb{C}(x)$.
$$\mathbb{G}_k=\left\{\begin{pmatrix}\lambda&d\\0&\lambda^{-1}\end{pmatrix}:\quad \lambda\text{  is a $k$-root of the unity, }d\in\mathbb{C}\right\},$$
this group is disconnected and its identity component is abelian.

\item[I3] $\mathbb{C}^*$ when the algorithm provides two
non-algebraic solutions.
$$\mathbb{C}^*=\left\{\begin{pmatrix}c&0\\0&c^{-1}\end{pmatrix}:c\in \mathbb{C}^*\right\},$$
this group is connected and abelian.

\item[I4] $\mathbb{C}^{+}$ when the algorithm provides one
rational solution and the second solution is not algebraic.
$$\mathbb{C}^{+}=\left\{\begin{pmatrix}1&d\\0&1\end{pmatrix}:d\in \mathbb{C}\right\}, \quad {\xi}\in\mathbb{C}(x),$$
this group is connected and abelian.

\item[I5] $\mathbb{C}^*\ltimes\mathbb{C}^+$ when the algorithm only provides one
solution $\xi$ such that $\xi$ and its square are not rational
functions.
$$\mathbb{C}^*\ltimes\mathbb{C}^{+}=\left\{\begin{pmatrix}c&d\\0&c^{-1}\end{pmatrix}:c\in\mathbb{C}^*,d\in \mathbb{C}\right\}, \quad \xi\notin\mathbb{C}(x),\quad {\xi}^2\notin\mathbb{C}(x).$$
This group is connected and non-abelian.

\item[I6] $\mathrm{SL}(2,\mathbb{C})$ if the algorithm does not provide any
solution. This group is connected and non-abelian.
\end{description}
\end{remark}

\begin{remark}[Case 2]If the RLDE falls in case 2, then Kovacic's Algorithm can provide us one or two
solutions. This depends on $r$ as follows:
\begin{description}
\item[II1] if $r$ is given by
$$r={2\phi'+2\phi-\phi^2\over 4},$$
then there exist only one solution,

\item[II2] if $r$ is given by
$$r\neq{2\phi'+2\phi-\phi^2\over 4},$$
then there exists two solutions.
\item[II3] The identity component of the Galois group for this
case is abelian.
\end{description}
\end{remark}

\begin{remark}[Case 3]If the RLDE falls in case 3, then its Galois group is given by one of
the following groups:
\begin{description}
\item[III1] {\bf Tetrahedral group} when $\omega$ is obtained with $n=4.$ This group of order 24 is generated by
$$
\begin{pmatrix}
e^{\frac{k\pi i}{3}} & 0\\
0 & e^{-\frac{k\pi i}{3}}
\end{pmatrix}
, \quad\frac{1}{3}\left(  2e^{\frac{k\pi i}{3}}-1\right)
\begin{pmatrix}
1 & 1\\
2 & -1
\end{pmatrix},\quad k\in \mathbb{Z}.$$
\item[III2] {\bf Octahedral group} when $\omega$ is obtained with $n=6.$ This group of order 48 is generated by
$$
\begin{pmatrix}
e^{\frac{k\pi i}{4}} & 0\\
0 & e^{-\frac{k\pi i}{4}}
\end{pmatrix}, \quad \frac{1}{2}e^{\frac{k\pi i}{4}}\left(  e^{\frac{k\pi
i}{2}}+1\right)
\begin{pmatrix}
1 & 1\\
1 & -1
\end{pmatrix},\quad k\in \mathbb{Z}.$$

\item[III3]{\bf Icosahedral group} when $\omega$ is obtained with $n=12.$ This group of order 120 is generated by
$$
\begin{pmatrix}
e^{\frac{k\pi i}{5}} & 0\\
0 & e^{-\frac{k\pi i}{5}}
\end{pmatrix}
,\quad
\begin{pmatrix}
\phi & \psi\\
\psi & -\phi
\end{pmatrix},\quad k\in \mathbb{Z},$$ being $\phi$ and $\psi$ defined as
$$\phi=\frac{1}{5}\left(  e^{\frac{3k\pi
i}{5}}-e^{\frac{2k\pi i}{5} }+4e^{\frac{k\pi i}{5}}-2\right),
\quad \psi=\frac{1}{5}\left( e^{\frac{3k\pi i}{5}}+3e^{\frac{2k\pi
i}{5}}-2e^{\frac{k\pi i}{5}}+1\right)$$

\item[III4] The identity component of the Galois group for this
case is abelian.
\end{description}
\end{remark}

\section{Applications}

\subsection{The collinear restricted elliptic three-body problem}
Let two primaries of mass $1-\mu$ and $\mu$ move along the $x$-axis,
its positions being $x_1=-\mu r$, $x_2=(1-\mu) r$ where
$\mu=m_2/(m_1+m_2)$. Suppose the primaries perform an elliptic collision motion
\begin{eqnarray*}
  r &=&1-\cos E \\
  t &=& E-\sin E
\end{eqnarray*}
where  $E$ is the elliptic anomaly and we choose units of time and length such that
the maximum distance between  the primaries is unit and the mean motion is one.
 The equations of motion of
a massless particle in a fixed plane containing the line of the primaries is
\begin{eqnarray}
    \ddot{x}&=&
  -\frac{(1-\mu)(x-x_1)}{|x-x_1|^3}-\frac{\mu(x-x_2)}{|x-x_2|^3}\nonumber\\
&=& -\frac{(1-\mu)(x+\mu r)}{|x+\mu r|^3}-
\frac{\mu(x-(1-\mu)r)}{|x-(1-\mu) r|^3}\label{massless}
\end{eqnarray}
where  its position is $x\in\C$. System (\ref{massless}) is two
degrees of freedom time--dependent Hamiltonian system. Some
general results are known for time--dependent, one degree of
freedom (see for example \cite{ac}). For the present
 we take an ad-hoc procedure:
We will perform several changes of variables in order to obtain the desired
form of equations of motion. Firstly, perform a change to pulsating coordinates
\begin{equation}\label{pulsating}
    x = r \xi
\end{equation}
then (\ref{pulsating}) transforms into
\begin{equation}
r\ddot{\xi}+2\dot{r}\dot{\xi}+\ddot{r}\xi=\frac{1}{r^2}\left[
-\frac{(1-\mu)(\xi+\mu )}{|\xi+\mu |^3}-
\frac{\mu(\xi-1+\mu)}{|\xi-1+\mu|^3}\right].
\end{equation}
Using the elliptic anomaly  $E$ as independent variable,
$$
\frac{d}{dt}=\frac{1}{r}\frac{d}{dE}
$$
yields
\begin{eqnarray*}
r \frac{1}{r}\frac{d}{dE}\left(\frac{1}{r}\frac{d\xi}{dE}\right) +2
\frac{1}{r}\frac{dr}{dE}\frac{1}{r}\frac{d\xi}{dE}&=&
\frac{1}{r^2}\left[\xi -\frac{(1-\mu)(\xi+\mu )}{|\xi+\mu |^3}-
\frac{\mu(\xi-1+\mu)}{|\xi-1+\mu|^3}\right],\\
\frac{d}{dE}\left(\frac{1}{r}\frac{d\xi}{dE}\right)+\frac{2\sin
E}{r^2}\frac{d\xi}{dE}&=&\frac{1}{r^2}\nabla \Omega(\xi)\\
\end{eqnarray*}
where the potential function is
\begin{equation}\label{Omega}
\Omega(\xi)=\frac{1}{2}|\xi|^2+\frac{1-\mu}{|\xi+\mu|}+\frac{\mu}{|\xi-1+\mu|}.
\end{equation}
Developing the left hand side of the previous ode we obtain
\begin{eqnarray*}
-\frac{\sin E}{r^2}\frac{d\xi}{dE}+\frac{1}{r}\frac{d^2\xi}{d
E^2}+\frac{2\sin E}{r^2}\frac{d\xi}{dE}&=&\frac{1}{r^2}\nabla \Omega(\xi)\\
\frac{1}{r}\frac{d^2\xi}{dE^2}+\frac{\sin
E}{r^2}\frac{d\xi}{dE}&=&\frac{1}{r^2}\nabla\Omega(\xi)\\
\frac{d^2\xi}{dE^2}+\frac{\sin E}{r}\frac{d\xi}{dE}&=&
\frac{1}{r}\nabla\Omega(\xi).
\end{eqnarray*}
In summary,
\begin{equation}\label{main}
\frac{d^2\xi}{dE^2}+\frac{\sin E}{1-\cos E}\,\frac{d\xi}{dE}=
\frac{1}{1-\cos E}\nabla\Omega(\xi)
\end{equation}
The equation (\ref{main}) can be analytically extended  to the whole
complex $E$-plane except for singularities at the point on the real axis $E=\pm\pi,\pm 2\pi,\ldots$,
and also has singularities due to collisions with the binaries
$\xi=-\mu,1-\mu$.

The critical points of (\ref{Omega}) are given by the
classical Eulerian and
Lagrangian points satisfying $\nabla \Omega(\xi_{L_i})=0$, $i=1,2\ldots, 5$.
Let $B$ denote the Hessian
$$
B=\left(\begin{array}{cc} \Omega_{11} &\Omega_{12}\\
\Omega_{21}& \Omega_{22}\end{array}\right)
$$
evaluated at any of the points $L_i$. The linearization
of(\ref{main}) at $L_i$ is
\begin{equation}\label{main-linear}
\frac{d^2\xi}{dE^2}+\frac{\sin E}{1-\cos E}\,\frac{d\xi}{dE}=
\frac{B\xi}{1-\cos E}.
\end{equation}
The above procedure can be seen as the linearization of the lifted system
\begin{eqnarray*}
\frac{dE}{ds} &=& 1,\\
\frac{d\xi}{ds} &=& v,\\
 \frac{d v}{ds} &=&-\frac{\sin E}{1-\cos E}\,\frac{d\xi}{dE}-
\frac{B\xi}{1-\cos E}
  \end{eqnarray*}
where $E$ is considered $\mbox{mod}\, 2\pi$,
along any of the periodic orbits $\xi=\xi_{L_i}$, $i=1,2,3$, $E=s.$

The following properties of matrix $B$ are well known (for details see \cite{szebe}):
For collinear configurations  $L_1,L_2,L_3$, $ B=\mbox{diag}(\kappa_1,\kappa_2)$,
 with $\kappa_1>0$ and $\kappa_2<0$ for all values of the mass
 parameter $\mu$, the exact values depend on the root of Euler's
 quintic equations.

In this case, the variational equations (\ref{main-linear}) split
\begin{equation}\label{split}
\frac{d^2\xi_j}{dE^2}+\frac{\sin E}{1-\cos E}\,\frac{d\xi_j}{dE}=
\frac{\kappa_j\xi_j}{1-\cos E},\quad \textrm{where} \; j=1,2.
\end{equation}

\begin{theorem}\label{homogeneous} For
the collinear elliptic restricted three body problem in the plane (\ref{main})
let
$$
\Omega(\xi)=\frac{1}{2}|\xi|^2+\frac{1-\mu}{|\xi+\mu|}+\frac{\mu}{|\xi-1+\mu|}
$$
and let $J$ the set of exceptional mass parameters $\mu\in (0,1)$ such that:
\begin{enumerate}
\item[{\rm (i)}] $\xi_1^{*}(\mu)$, $*=1,2,3$ belongs to any of the three solution curves
of  the equation
defining the  collinear configurations
$$
\frac{\partial\Omega}{\partial \xi_1}(\xi_1^*(\mu),0)=0;
$$
$-\infty<\xi_1^{1}(\mu)<-\mu$, \quad $-\mu< \xi_1^{2}(\mu)<1-\mu$, \quad $1-\mu <\xi_1^{3}(\mu)<\infty$;
\item[{\rm (ii)}] any of the coefficients
$$
\kappa_1^*(\mu)=\frac{\partial^2\Omega}{\partial\xi_1^2}(\xi_1^*(\mu),0),\qquad
\kappa_2^*(\mu)=\frac{\partial^2\Omega}{\partial\xi_2^2}(\xi_1^*(\mu),0)
$$
satisfy
\begin{equation}\label{kappas}
    \kappa_j^*= \frac{n(n+1)}{2}
\end{equation}
where $n$ is an integer.
\end{enumerate}
Then if $\mu\not\in J$, the problem is not integrable.
\end{theorem}
\proof

The procedure is to algebrize the variational equations
(\ref{split}) and then  apply Kovacic's algorithm. We start
considering the variational equation,
$$
\frac{d^2\xi_j}{dE^2} +\frac{\sin E}{1-\cos E}\frac{d\xi_j}{dE}-\frac{\kappa_j\: \xi_j}{1-\cos E}=0 \quad \textrm{where}
\; j=1,2, $$ which is transformed in the differential equation
\begin{equation}\label{marjoa2}
\frac{d^2y}{dE^2}=\phi(E)\:y(E)
\end{equation}
where
$$
\phi(E)=\frac{\cos E -1 + 4\kappa_j}{4(1-\cos E)},
\quad \xi_j(E)={y(E)\over \sqrt{1-\cos E }}.
$$
Now, by theorem \ref{pr2}, the equation \eqref{marjoa2} is ready
to be algebrized.  The Hamiltonian change of variable is
$\tau=\tau(E)=\cos E$, where $\dot\tau=-\sin E$,
$(\dot\tau)^2=\sin^2 E=1-\cos^2 E$ so that $$\alpha=1-\tau^2
\quad \textrm{and}\quad f=\frac{\tau-1+4\kappa_j}{4(1-\tau)}.$$
The algebraized equation is
\begin{equation}\label{marjoa3}
\frac{d^2y(\tau)}{d\tau^2}-{\tau\over 1-\tau^2}{dy(\tau)\over
d\tau }+{\tau-1+4\kappa_j\over 4(-1+\tau)(1-\tau^2)}y(\tau)=0,
\end{equation}
and the points $1,$ $-1$ and $\infty$ are regular singularities.
To apply Kovacic's algorithm, see section~\ref{four-cases}, we use the
RLDE
\begin{equation}\label{marjoa4}
\frac{d^2\eta}{d\tau^2}=r(\tau)\eta,\qquad r(\tau)=\frac{4\kappa_j
\tau+4\kappa_j-3}{4(1-\tau)^2(1+\tau)^2},\quad y(\tau)={\eta\over\sqrt[4]{1-\tau^2}}
\end{equation}
with $\kappa_j\neq 0$, because $\kappa_1>0$ and $\kappa_2<0$. We can
see that $\Upsilon=\{-1,1,\infty\}$ and that the equation
\eqref{marjoa4} could fall in any of four cases of Kovacic's
algorithm, now expanding $r(\tau)$ in partial fractions we have
that
$$r(\tau)={8\kappa_j - 3\over 16(1-\tau)^2} + {4\kappa_j -3 \over 16·(1-\tau)}
- {3\over 16·(1+\tau)^2} + {4\kappa_j - 3\over 16·(1+\tau)}.$$
 We start analyzing the case one. The equation
\eqref{marjoa4} satisfy the conditions $\{c_2,\infty_1\}$, because
$\circ r_1=\circ r_{-1}=2$ and $\circ r_{\infty}=3$, obtaining the
expressions
$$\begin{array}{l}
[\sqrt{r}]_{-1}=[\sqrt{r}]_{1}=[\sqrt{r}]_{\infty}=\alpha^+_{\infty}=0,\quad
\alpha^-_{\infty}=1,
\\
\\
 \alpha^+_{-1}=\frac34,\quad
 \alpha^-_{-1}=\frac14,\quad\alpha^{\pm}_{1}={2\pm\sqrt{8\kappa_j + 1}\over 4}.\end{array}$$ By step two, $D=\mathbb{Z}_+$ and $\kappa_j$ has the following
 possibilities:
$$ \kappa_j=(n+1)(2n+3),\quad \kappa_j=(n+1)(2n+1),\quad \kappa_j=n(2n+1),\quad \kappa_j=n(2n-1)$$
which are equivalents to $\kappa_j=n(n+1)/2$. For each $n$ we can
construct $\omega$ and by step three there exists a monic
polynomial of degree $n$ in which each solution of the
differential equation \eqref{marjoa4} is given for all
$n\in\mathbb{Z}_+$.

Following the case two, we expect to find different values of
$\kappa_j$ that the presented in case one, so that the equation
\eqref{marjoa4} satisfy the conditions $\{c_2,\infty_1\}$, because
$\circ r_1=\circ r_{-1}=2$ and $\circ r_{\infty}=3$, obtaining the
expressions
$$
E_1=\left\{2,2-\sqrt{1+8\kappa_1},2+\sqrt{1+8\kappa_1}\right\},\quad
E_{-1}=\{1,2,3\},\quad E_{\infty}=\{0,2,4\}.$$ By step two,
$D=\mathbb{Z}_+$ and $\kappa_j$ is again equivalent to
$\kappa_j={n(n+1)\over 2},$ so that we discard the case two.

Finally, following the case 3, we expect to find different values
of $\kappa_j$ that the presented in case one, but again appear the
expression $\sqrt{1+8\kappa_j}$, which replaced in $E_c$ and
$E_\infty$ give us again an equivalent expression to
$\kappa_j={n(n+1)\over 2}.$ This means that the differential
equation \eqref{marjoa4} is contained in the Borel group when
$\kappa_j={n(n+1)\over 2},$ and it is $\mathrm{SL}(2,\mathbb{C})$
when $\kappa_j\neq{n(n+1)\over 2}.$ Therefore, by remark
\ref{rkov2}, the Galois group is virtually abelian for
$\kappa_j={n(n+1)\over 2}$ and unsolvable for
$\kappa_j\neq{n(n+1)\over 2}$.
\qed

An alternative proof based on Kimura's approach is given in the  Appendix~A.1, mainly
for contrasting both techniques.
\bigskip

It is interesting to investigate the exceptional values of the mass parameter $\mu\in J$ such that any of the $\kappa_1$, $\kappa_2$ satisfy the condition (\ref{kappas}). Since $\kappa_2$ is negative, there are no exceptional values since $\frac{n(n+1)}{2}\geq 0$ for all integers
$n$. Shown in Figure \ref{gr} are the curves  $\kappa_{1,2}^{*}(\mu)$, for $*=1,2$ ($\kappa_1$, $\kappa_2$ are symmetrical with respect to
$\mu=1/2$, so we just consider $L_1$ and $L_2$).
For $\mu=0$, the non--integrability test fails since then both $\kappa_1^{1}(0)=2\cdot \frac{3}{2}=3$
and $\kappa_1^{2}(0)=3\cdot \frac{4}{2}=6$ (can be verified analytically). This is consistent with the fact that for $\mu=0$
system
is just a Kepler problem. The exceptional values $\mu_{L_1}$ satisfying
 $\kappa_1(\mu_{L_1})=3\cdot\frac{4}{2}=6$ and
$\mu_{L_2}<0.5$ satisfying $\kappa_1(\mu_{L_2})=3\cdot\frac{4}{2}=6$ are not satisfied {\em simultaneously}
for the same value of the mass parameter, i.e.
$\mu_{L_1}\neq \mu_{L_2}$, thus for {\em some of the reference orbits} $L_1$ or $L_2$,  the system does not posses an integral in a neighborhood
of that orbit, although the theorem does not discard the existence of an additional integral locally defined.
\begin{figure}
\centering
\textbf{Sorry, the graphics are not available.}
  \caption{(a) The curves $\kappa_1^{1}(\mu)$ (up) and $\kappa_2^{1}(\mu)$ (bottom).
  (b) The curves $\kappa_1^{2}(\mu)$ (up) and $\kappa_2^{2}(\mu)$ (bottom).}\label{gr}
\end{figure}

\subsection{Homogeneous potential of degree $-1$}
We consider a general application to a two degrees of freedom simple hamiltonian system
with homogeneous potential of degree $-1$
$$
 H =\frac{1}{2}(p_x^2+p_y^2)-U(x,y),
$$
We suppose that
$U(x,y)$ is defined  and is positive for all $(x,y)\in\R^2$, except the origin.

The Hamiltonian un polar coordinates becomes
$$
H=\frac{1}{2}\left(p_r^2+\frac{p_\theta^2}{r^2}\right)-\frac{1}{r}U(\theta).
$$
McGehee's blow up is achieved taking coordinates $v=  r^{-1/2} p_r$, $u = r^{-1/2}p_\theta$
and rescaled time $dt=r^{3/2}d\tau$.
The equations of motion take the form
 \begin{eqnarray*}
r'&=&rv, \nonumber\\
 v'&=&\frac{1}{2}v^2+u^2-U(\theta), \label{McGehee}\\
\theta' &=& u, \nonumber\\
u'& =& -\frac{1}{2}vu+ U'(\theta).\nonumber
\end{eqnarray*}
Where the prime in the left hand side denotes derivatives with respect to $\tau$ and
$U'(\theta)$
denotes derivative with respect to its argument, which causes no confusion.
System (\ref{McGehee}) leaves invariant  the energy surface
\begin{equation}\label{energy-surface}
E_h=\{(r,\theta, u,v)\mid r>0,\quad \frac{1}{2}(u^2+v^2)=U(\theta)+rh\},
\end{equation}
which can be extended invariantly up to its boundary, the collision manifold
\begin{equation}\label{collision-manifold}
\Lambda=\{(r,\theta, u,v)\mid r=0,\quad \frac{1}{2}(u^2+v^2)=U(\theta)\}.
\end{equation}

Since  $U(\theta)$ is periodic, by the mean value theorem, there exists $\theta_c$ such that $U'(\theta_c)=0$.
Let $v_c=2U(\theta_c)$. Then for $h<0$ there exists an ejection--collision homothetic orbit given explicitly
by $\theta=\theta_c$, $u=0$ and
\begin{eqnarray*}
r_h(\tau) &=& -\frac{v_c^2}{2h}\sech^2(v_c\tau/2), \\
v_h(\tau) &=& -v_c \tanh(v_c\tau/2) .
\end{eqnarray*}
The variational equations along the homothetic orbit are
\begin{eqnarray*}
\delta r' &=& v_{h}\delta r + r_{h}\delta v,\\
\delta v'  &=& v_{h}\delta v,\\
\delta\theta' &=& \delta u,\\
\delta u' &=& -\frac{1}{2}v_{h}\delta u + U''(\theta _c)\delta\theta
\end{eqnarray*}
The last two equations are decouple and constitute the normal variational equations. They
can be expressed with respect to the scaled time  $s=v_c\tau/2$, that will still be denoted
by primes,
\begin{equation}\label{var}
\delta\theta ''(s) -\tanh(s)\delta\theta'(s)-\omega^2\theta(s) =0
\end{equation}
where
$$
\omega^2=\frac{2 U''(\theta_c)}{U(\theta_c)}.
$$
($\omega$ can be imaginary).

\begin{remark}
McGehee's equations (\ref{McGehee}) are hamiltonian with respect to the symplectic form
$\alpha=2dv\wedge dr^{1/2}+d(r^{1/2} u)\wedge d\theta$ obtained by pullback of the canonical form
$dp_x\wedge dx+dp_y\wedge dy$ under  McGehee transformation. Therefore Morales--Ramis applies to this case.

\end{remark}
\begin{theorem}\label{ours}
Let the Hamiltonian of a system be
$$
H =\frac{1}{2}(p_x^2+p_y^2)-U(x,y)
$$
with $U(x,y)>0$ homogeneous of degree $-1$, defined for all
$(x,y)\neq(0,0)$. Let
$U'(\theta_c)=0$ and
  Let $\omega^2=\frac{2 U''(\theta_c)}{U(\theta_c)}$, then if
\begin{equation}\label{condn}
\omega^2 \neq n(n+1) \quad\mbox{$n$ being an integer},
\end{equation}
then on a fixed negative energy level the system has no meromorphic integral
in a neighborhood of the homothetic solution defined by $\theta=\theta_c$.
\end{theorem}
\proof

Consider the variational equation
\eqref{var}
$${d^2z\over dt^2}-\tanh (t){dz\over dt}-\omega^2z=0,$$
which is transformed into the differential equation
\begin{equation}\label{marjoa21}
\frac{d^2y(t)}{dt^2}=\phi(t)y(t),\, \phi(t)={\frac
{\cosh^2(t)+4\omega^{2}\cosh^2(t)-3}{4 \cosh^2 (t)}} ,\,
z(t)=y(t)\sqrt{\cosh(t)}.
\end{equation}
Now, by theorem \ref{pr2}, the equation \eqref{marjoa21} is ready
to be algebrized.  The Hamiltonian change of variable is
$\tau=\tau(t)=\cosh (t)$, where $\dot\tau=\sinh (t)$,
$(\dot\tau)^2=\sinh^2 (t)=-1+\cosh^2(t)$ so that
$$\alpha=-1+\tau^2 \quad \textrm{and}\quad
f=\frac{\tau-1+4\kappa_j}{4(1-\tau)}.$$ The algebraized equation
is
\begin{equation}\label{marjoa31}
\frac{d^2y(\tau)}{d\tau^2}-{\tau\over 1-\tau^2}{dy(\tau)\over
d\tau }-{(1+4\omega^2)\tau^2 -3\over 4\tau^2(\tau^2-1)}y(\tau)=0,
\end{equation}
and the points $0$, $1,$ $-1$ and $\infty$ are regular
singularities. To apply Kovacic's algorithm, see appendix A, we
use the RLDE
\begin{equation}\label{marjoa41}
\frac{d^2\eta}{d\tau^2}=r(\tau)\eta,\qquad
r(\tau)={4\omega^2\tau^4-(6+4\omega^2)\tau^2 +3\over
4\tau^2(\tau-1)^2(\tau+1)^2},\quad y(\tau)={\eta\over
\sqrt[4]{1-\tau^2}}
\end{equation}
with $\omega\neq 0$, because with $\omega=0$ the differential
equation can be solved easily. We can see that
$\Upsilon=\{0,-1,1,\infty\}$ and that the equation
\eqref{marjoa41} could fall in any of four cases of Kovacic's
algorithm, now expanding $r(\tau)$ in partial fractions we have
that
$$r(\tau)=- {3\over 16(\tau - 1)^2} + {8\omega^2 - 3\over 16·(\tau - 1)} - {3\over 16(\tau + 1)^2} + {3 - 8·\omega^2\over 16·(\tau + 1)} + {3\over 4\tau^2}.$$
 We start analyzing the case one. The equation
\eqref{marjoa41} satisfy the conditions $\{c_2,\infty_2\}$,
because $\circ r_1=\circ r_{-1}=1=\circ r_0=\circ r_{\infty}=2$.
Due to condition $\infty_2$, we need the Laurent series of
$r(\tau)$ around $\infty$, which corresponds to
$$r(\tau)=\omega^2\tau^2+\left(-\frac32+\omega^2\right)\tau^4+\left(\omega^2-\frac94\right)\tau^6+O(\tau^8)$$
obtaining the expressions
$$\begin{array}{l}
[\sqrt{r}]_{0}=[\sqrt{r}]_{-1}=[\sqrt{r}]_{1}=[\sqrt{r}]_{\infty}=0,\quad
\\
\\
\alpha^+_{1}=\alpha^+_{-1}=\frac34,\quad
\alpha^-_{1}=\alpha^-_{-1}=\frac14,\quad
\alpha^\pm_{\infty}={1\pm\sqrt{1+4\omega^2}\over 2}.
\end{array}$$ By step two, $D=\mathbb{Z}_+$
  and $\omega^2$ has the following  possibilities:
$$\begin{array}{l}
 \omega^2=(n+2)(n+3),\quad \omega^2={(2n+3)(2n+5)\over 4},\quad \omega^2=(n+1)(n+2),\\ \\ \omega^2=n(n+1),\quad\omega^2={(2n+1)(2n-1)\over 4},\quad\omega^2=n(n-1), \end{array}$$
discarding $\omega^2\notin\mathbb{Z}$ because the differential
equation has not Liouvillian solutions (the monic polynomial $P_n$
there is not exists), we take the rest of values for $\omega^2$
which are equivalents to $\omega^2=n(n+1)$. For each $n$ we can
construct $\omega$ and by step three there exists a monic
polynomial of degree $n$ in which each solution of the
differential equation \eqref{marjoa41} is given for all
$n\in\mathbb{Z}_+$.

Following the case two, we expect to find different values of
$\omega^2$ that the presented in case one, so that the equation
\eqref{marjoa41} satisfy the conditions $\{c_2,\infty_2\}$,
because $\circ r_0=\circ r_1=\circ r_{-1}=\circ r_{\infty}=2$,
obtaining the expressions
$$
E_{0}=\{-2,2,6\},\quad E_1=E_{-1}=\{1,2,3\},\quad
E_{\infty}=\left\{2,2-\sqrt{1+4\omega^2},2+\sqrt{1+4\omega^2}\right\}.$$
By step two, $D=\mathbb{Z}_+$ and we obtain again
$\omega^2=n(n+1),$ so that we discard the case two.

Finally, following the case 3, we expect to find different values
of $\omega^2$ that the presented in case one, but again appear the
expression $\sqrt{1+4\omega^2}$, which replaced in $E_c$ and
$E_\infty$ give us again $\omega^2=n(n+1).$ This means that the
differential equation \eqref{marjoa41} is contained in the Borel
group when $\omega^2=n(n+1),$ and it is
$\mathrm{SL}(2,\mathbb{C})$ when $\omega^2\neq n(n+1).$ Therefore,
by remark \ref{rkov2}, the Galois group is virtually abelian for
$\omega^2=n(n+1)$ and unsolvable for $\omega^2\neq n(n+1)$.
\qed

An alternative proof based on Kimura's approach in given in Appendix A.2.

\begin{remark}
Yoshida \cite{yoshida} gives sufficient conditions for the non--integrability
for Hamiltonian systems of two degrees of freedom $H=(p_1^2+p_2^2)/2+V(q_1,q_2)$
 with homogeneous potential $V$
of arbitrary integer degree $k$. He defines the ``integrability coefficient"
$
\lambda =\emph{Trace}(\emph{Hess\,} V(c_1,c_2))-(k-1)
$
where $(c_1,c_2)$ is a solution of the algebraic equation
\begin{equation}\label{trascendence}
(c_1,c_2)=\nabla V(c_1,c_2)
\end{equation}
and $\emph{Trace}(\emph{Hess\,} V(c_1,c_2))$ is the trace of the Hessian  matrix.
It is not difficult to show that the ``integrability coefficient" is related to
our parameter $\omega$ by $\omega^2/2=1-\lambda$.
Theorem \ref{ours} can be considered as equivalent to Yoshida's theorem in the
particular case $k=-1$, since condition (\ref{trascendence}) can be viewed
as the vanishing of the gradient of the restriction of $V|_{S^1}$
to the unit circle $S^1$, i.e. $V'(\theta)=0$.
\end{remark}

\begin{remark}
Vigo--Aguiar (cited in \cite{vigo}) and co-workers, have developed systematically the formulation
of Yoshida's result in  polar coordinates and used it to study two degrees of freedom polynomial potentials.
\end{remark}

\begin{remark}
The main difference of theorem (\ref{ours}) and previous work cited in the above remarks, is that we are
considering explicitly the variational equations along a singular ejection--collision orbit.
\end{remark}

In the following subsections we  apply the theory developed so far to some examples
of few body problems. The main interest is to test the non--integrability given by
theorem~\ref{ours} in concrete examples having singularities. For simplicity,  the
equations of motion in the examples are recast in McGehee's form (\ref{McGehee}).

\subsubsection{The rectangular 4 body problem}
Four unit masses are at the vertices of a rectangle
with initial conditions (position and velocity) symmetrical with
respect to the axes in such a way that the rectangular
configuration of the particles is preserved. See Figure~\ref{rectangular}.
\begin{figure}
\centering \textbf{Sorry, the graphic is not available.}
\caption{Rectangular $4$--body problem}\label{rectangular}
\end{figure}
Let  $x$, $y$ be the base and height of
the rectangle with  the center of mass at the origin, $p_x=\dot{x}$, $p_y=\dot{y}$ conjugate momenta.
The Hamiltonian is
$$
H=\frac{1}{2}(p_x^2+p_y^2)-\frac{1}{x}-\frac{1}{y}-\frac{1}{\sqrt{x^2+
y^2}}
$$
Taking polar--like coordinates
$x = r\cos\theta$, $y=r\sin\theta$
the equations of motion are of the type (\ref{McGehee}) with
$$
U(\theta)=\frac{1}{\cos \theta }+\frac{1}{\sin \theta }+1.
$$
The unique homothetic orbit
corresponds to $\theta_c=\pi/4$. A simple computation shows that
$$
\omega^2=\frac{12\sqrt{2}}{1+2\sqrt{2}}
$$
then from theorem (\ref{ours}) it follows trivially,
\begin{theorem}
The rectangular four body problem is not integrable with meromorphic first integrals.
\end{theorem}

\subsubsection{The anisotropic Kepler problem}
The hamiltonian of the anisotropic Kepler
$$
H=\frac{1}{2}(p_x^2+p_y^2)-\frac{1}{\sqrt{x^2+\mu y^2}}
$$
depends on the  parameter of anisotropy which can be restricted to $\mu\in [0,1]$. For $\mu=0$ and $\mu=1$
it is integrable. Using polar coordinates $x=r\cos\theta$, $y=r\sin\theta$ McGehee's equation
are obtained (\ref{McGehee}) with
$$
U(\theta)=\frac{1}{\sqrt{\cos^2\theta+\mu\sin^2\theta}}.
$$
Homothetic orbits correspond to minima at  $\theta=0,\pi$ and maxima at $\theta=\pi/2,3\pi/2$;
then for minima
$$
\omega^2=2(1-\mu.)
$$
According to the non-integrability theorem (\ref{ours}), the anisotropic Kepler problem is not integrable with meromorphic integrals
except when $\mu=1-\ell(\ell+1)/2$. This leaves only the integrable  cases $\mu=0,1$. Figure~\ref{anis} shows the Poincar\'e maps
associated to the section $\theta=0$ for some values of the mass parameter.
\begin{figure}
\centering
\textbf{Sorry, the graphics are not availables.}
\caption{Poincar\'e maps for the anisotropic Kepler problem (a)
$\mu=1$; (b) $\mu=0.9$; (c) $\mu=0.85$; (d) $\mu=0$. Integrable
cases $\mu=0,1$ were calculated analitically. The main role of two
hyperbolic orbits is evident.}\label{anis}
\end{figure}

\subsubsection{Two uncoupled Kepler problems}
Consider two uncoupled Kepler problems on the line with Hamiltonian
$$
H=\frac{1}{2}(p_x^2+p_y^2)-\frac{1}{x}-\frac{\mu}{y},
$$
which models, for example, two binaries on the line far apart so  the interaction between them can be neglected.
In one binary the particles have the same mass taken as unit, and on the other binary $\mu$ represents its total mass.
This problem is evidently integrable for all values of $\mu$. Using polar coordinates $x=r\cos\theta$, $y=r\sin\theta$,
the potential becomes
$$
U(\theta)= \frac{1}{\cos\theta}+\frac{\mu}{\sin\theta}.
$$
The unique critical point corresponds to the homothetic orbit $\theta_c=\arctan(\mu^{1/3})$. One easily computes
$$
\omega^2=6
$$
which does not depend on $\mu$. Thus the non-integrability test (\ref{ours}) fails, accordingly since the problem is completely
integrable.

\section{Open questions and final remarks}
Kovacic algorithm and Kimura's table give the same non--integrability results in the specific examples studied in this
paper. We recover Yoshida's non-integrability in the case of simple mechanical system with two degrees of freedom an homogeneous
potential of degre $-1$, by first performing McGehee's the blow up. Yoshida's approach and ours are not entirely equivalent though, since here we are considering specifically an ejection--collision orbit exhibiting singularities  in the original coordinates. This singularity is substituted by an invariant manifold and the singular orbit now connects two singular points in the collision manifold.
The  general setting, as stated by Morales \cite{mo1}, of adding singularities to the original Riemann surface is here needed in order
to apply the theory. To our knowledge only one such example is known where this kind of generality is needed, the Bianchi IX cosmologial model, and has been discussed in great detail by Morales and Ramis in \cite{moralesramisII}. Based in this situation  we pose the following open problem:

Consider a Hamiltonian system on a fixed energy level $M_h$ with with an invariant submanifold $\Lambda$  on its boundary. The flow
preserves the natural volume form on $M_h$ but not necessarily on $\overline{M_h}$. Let  $\gamma$ be a heteroclinic (homoclinic) orbit connecting critical points on $\Lambda$ but not completely contained in $\Lambda$. We ask: which are the class of integrals that are dismissed by Morales--Ramis theory? For example, if such an alleged first integral is continuous up to $\Lambda$ then by invariance it has to be constant on $\Lambda$ (following Abraham--Marsden,  we call such integrals {\em extendable}), this
is clearly a strong restriction. If such an integral has
poles on $\Lambda$ are the critical points  on $\Lambda$  necessarily one of them?
Another open question is to investigate how is transversality of stable and unstable manifolds
along $\gamma$ related to the solvability of the differential Galois group. As a reference, Yagasaki \cite{yagasaki} gives an answer to this question in the case of $\Lambda=\{p\}$ a critical point an the extended flow $\overline{M_{h}}$ is still volume preserving (we just add a critical point).

\appendix
\section{Kimura's Theorem}


The hypergeometric (or Riemann) equation is the more general second order linear
differential equation over the Riemann sphere with three regular singular
singularities. If we place the singularities at $x = 0, 1, \infty$ it is given by
\begin{eqnarray}\label{hypergeometric1}
&& \frac{d^2\xi}{dx^2}+ \left(\frac{1-\alpha-\alpha'}{x} + \frac{1-\gamma-\gamma'}{x-1}\right)\frac{d\xi}{dx}\\
&& \qquad\qquad\qquad \qquad  + \left(\frac{\alpha\alpha'}{x^2} +  \frac{\gamma\gamma'}{(x-1)^2}
+ \frac{\beta\beta'-\alpha\alpha'\gamma\gamma'}{x(x-1)}\right) \xi = 0, \nonumber
\end{eqnarray}
where $(\alpha , \alpha ')$, $(\gamma , \gamma ')$, $(\beta , \beta ')$
are the exponents at the singular points and must
satisfy the Fuchs relation $\alpha + \alpha' + \gamma + \gamma'+\beta + \beta'= 1$.

Now, we will briefly describe here the theorem of Kimura that gives necessary and sufficient
conditions for the hypergeometric equation to have integrability. Let be
$\hat{\lambda} = \alpha -\alpha'$, $\hat{\mu} = \beta-\beta'$ and $\hat{\nu} = \nu -\nu'$.

\begin{theorem}[\cite{kimura}]\label{kimurath}
The identity component of the Galois group of the hypergeometric equation (\ref{hypergeometric1}) is solvable
if and only if, either
\begin{enumerate}
\item[(i)] At least one of the four numbers $\hat{\lambda}+\hat{\mu}+\hat{\nu}$,
$-\hat{\lambda}+\hat{\mu}+\hat{\nu}$, $\hat{\lambda}-\hat{\mu}+\hat{\nu}$,
$\hat{\lambda}+\hat{\mu}-\hat{\nu}$ is an odd integer, or
\item[(ii)] The numbers $\hat{\lambda}$ or $-\hat{\lambda}$, $\hat{\mu}$
or $-\hat{\mu}$ and $\hat{\nu}$ or $-\hat{\nu}$ belong (in an arbitrary order) to
some of the following fifteen families
\end{enumerate}
$$
\begin{array}{|c|c|c|c|c|}\hline
1 & 1/2+l & 1/2+m & \mbox{arbitrary complex number} &\\ \hline
2 & 1/2+l & 1/3+m & 1/3+q &\\ \hline
3 & 2/3+l & 1/3+m & 1/3+q & l+m+q\mbox{ even}\\ \hline
4 & 1/2+l & 1/3+ m & 1/4+q & \\ \hline
5 & 2/3+l & 1/4+m & 1/4+q & l+m+q\mbox{ even}\\ \hline
6 & 1/2+l & 1/3+m & 1/5+q & \\ \hline
7 & 2/5+l & 1/3+m & 1/3+q & l+m+q\mbox{ even}\\ \hline
8 & 2/3+l & 1/5+m & 1/5+q & l+m+q\mbox{ even}\\ \hline
9 & 1/2+l & 2/5+m & 1/5+q & l+m+q\mbox{ even}\\ \hline
10 & 3/5+l & 1/3+m & 1/5+q & l+m+q\mbox{ even}\\ \hline
11 &2/5+l & 2/5+m & 2/5+q & l+m+q\mbox{ even}\\ \hline
12 &2/3+l & 1/3+m & 1/5+q & l+m+q\mbox{ even}\\ \hline
13 & 4/5+l & 1/5+m & 1/5+q & l+m+q\mbox{ even}\\ \hline
14 & 1/2+l &2/5+m & 1/3+q & l+m+q\mbox{ even}\\ \hline
15 & 3/5+l & 2/5+m & 1/3+q & l+m+q\mbox{ even}\\ \hline
\end{array}
$$
Here $n,m,q$ are integers.
\end{theorem}

\section{Alternative proof of theorem \ref{homogeneous}\label{kimura-homo}}
The change  of independent variable
\begin{equation}\label{reduction}
    z=\frac{1}{2}\left(\cos E+1\right).
\end{equation}
reduces the variational equations (\ref{split})  to the rational form
\begin{equation}\label{rational}
\frac{d^2\xi_j}{dz^2}+\left(\frac{3/2}{z}+\frac{1/2}{z-1}\right)\,\frac{d\xi_j}{dz}
+
\frac{\kappa_j}{2} \left(\frac{1}{z(z-1)}-\frac{1}{(z-1)^2}\right)\xi_j=0.
\end{equation}
By making the (non unique) choice of constants
\begin{eqnarray*}
&& \alpha'=\beta=0,\quad \alpha=-\frac{1}{2}, \quad \beta'=1,\\
&& \gamma= \frac{1}{4}\left(1+\sqrt{1+8\kappa_j}\; \right),
\qquad \gamma'= \frac{1}{4}\left(1-\sqrt{1+8\kappa_j}\;\right)
\end{eqnarray*}
the equations reduces to the hypergeometric equation of the form given in
(\ref{hypergeometric1}):
\begin{equation}\label{hypergeometric}
\frac{d^2\xi_j}{dz^2}+\left(\frac{1-\alpha-\alpha'}{z}+\frac{1-\gamma-\gamma'}{z-1}\right)\frac{d\xi_j}{dz}+
\left(\frac{\alpha\alpha'}{z^2}+\frac{\gamma\gamma'}{(z-1)^2}+
\frac{\beta\beta'-\alpha\alpha'-\gamma\gamma'}{z(z-1)}\right)\xi_j=0.
\end{equation}

In order to verify the theorem \ref{kimurath}, we define the difference of exponents
$$
  \hat{\lambda} = \alpha-\alpha'=-1/2, \qquad
  \hat{\mu} =\beta-\beta'= -1, \qquad
  \hat{\nu} = \gamma-\gamma'=\frac{1}{2}\sqrt{1+8\kappa_j}\:.
$$
In order to verify condition (i) of Kimura's theorem, we compute the combinations
\begin{eqnarray*}
    \hat{\lambda}+\hat{\mu}+\hat{\nu} &=& \frac{1}{2}(-3+\sqrt{1+8\kappa_j} )\\
    -\hat{\lambda}+\hat{\mu}+\hat{\nu} &=& \frac{1}{2}(-1+\sqrt{1+8\kappa_j} ) \\
    \hat{\lambda}-\hat{\mu}+\hat{\nu} &=& \frac{1}{2}(1+\sqrt{1+8\kappa_j} ) \\
    \hat{\lambda}+\hat{\mu}-\hat{\nu} &=& -\frac{1}{2}(3+8\sqrt{1+8\kappa_j} )
  \end{eqnarray*}
For any of the above quantities to be an odd integer, then $\kappa_j$ must be of the form
\begin{equation}\label{(i)}
    \kappa_j= (n+1)(2n+3),\, (n+1) (2n+1),\, n (2n+1),\quad n\in \Z.
\end{equation}
In order to verify condition (ii) observe that the only
possibility is that $\hat{\mu}$ fits in the column of ``arbitrary
complex number" and $\hat{\lambda}$ of the form $1/2+m$, with
$m=-1$ an integer, therefore the parameters $\kappa_j$, $j=1,2$
must satisfy the condition $\sqrt{1+8\kappa_j}=1/2+\ell$, or
\begin{equation}\label{ell}
    \kappa_j=\frac{1}{2}\ell(\ell+1).
\end{equation}
But conditions (\ref{(i)}) are contained in condition  (\ref{ell}), to see this take $\ell= 2(n+1),2n+1, 2n$, respectively to recover (\ref{(i)}).
\qed

\begin{remark}
 We recovered condition (\ref{kappas})
\end{remark}

\section{Alternative proof of theorem \ref{ours}\label{kimura-ours}}
The change of dependent variable
$$
\delta\theta(s)=\cosh^{1/2}(s)y(s)
$$
reduces the equation to
$$
y''(s)-\frac{1}{4}(1+4\omega^2-3\sech^2(s))y(s)=0.
$$
A further change of independent variable $z=\sech^2(s)$ yields
$$
dz=-2\sech^2(s)\tanh(s)\,ds=-2z\sqrt{1-z}\,ds.
$$
Thus
$$
\frac{d^2y}{ds}=4z\sqrt{1-z}\frac{d}{dz}\left(z\sqrt{1-z}\frac{dy}{dz}\right).
$$
developing the second derivative
\begin{eqnarray*}
4z^2(1-z)y''(z) +\left( 4z(1-z)+4z^2 \sqrt{1-z} \frac{-1}{2\sqrt{1-z}}\right)y'(z)
-\frac{1}{4}(1+4\omega^2-3z)y(z)&=&0\\
4z^2(1-z)y''(z) +2z \left( 2(1-z)- z\right)y'(z)
-\frac{1}{4}(1+4\omega^2-3z)y(z)&=&0\\
4z^2(1-z)y''(z)+2z(2-3z)y'(z)-\frac{1}{4}(1+4\omega^2-3z)y(z)&=&0\\
z(1-z)y''(z)+\frac{1}{2}(2-3z)y'(z)-\frac{1}{16z}(1+4\omega^2-3z)y(z)&=&0\\
y''(z)+\frac{1-\frac{3}{2}z}{z(1-z)}y'(z)+\frac{-\frac{1}{16}-\frac{\omega^2}{4}+\frac{3}{16}z}{z^2(1-z)}y(z)&=&0.
\end{eqnarray*}
Expanding in  partial fractions we finally get
\begin{equation}\label{hyper-2dof}
    y''(z)+\left(\frac{1}{z}+\frac{1/2}{z-1}\right)y'(z)+\left(\frac{-1/16-\omega^2/4}{z^2}+\frac{-1/8+\omega^2/4}{z(z-1)}\right)y(z)=0.
\end{equation}
which is a Riemman equation.
Comparing with (\ref{hypergeometric}) a convenient choice of parameters is $\gamma'=0$ and
$$\displaylines{
\alpha=-\frac{1}{4}\sqrt{1+4\omega^2},\quad \alpha'=\frac{1}{4}\sqrt{1+4\omega^2}\cr
\beta=\frac{1}{4},\quad \beta'=-\frac{3}{4},\quad\gamma=\frac{1}{2}.}
$$
The exponent differences are
\begin{eqnarray*}
  \hat{\lambda} &=&\alpha-\alpha'=-\frac{1}{2}\sqrt{1+4 \omega ^2} \\
  \hat{\mu} &=& 1 \\
  \hat{\nu} &=& \frac{1}{2}
\end{eqnarray*}
Condition (i) of Kimura's theorem is satisfied whenever any of the four combinations indicated there is an odd integer;
thus
\begin{equation}\label{appendix-(i)}
\frac{\omega^2}{2}=n(2n-1), n(2n+1),\quad \mbox{$n$ being an integer}
\end{equation}
To verify condition (ii) of Kimura's table, notice that $\hat{\mu}=-1$
is not of  any of the forms of the columns
except for the first case: We can take $\hat{\mu}$ as an ``arbitrary complex number"
and $\hat{\nu}=1/2$ of the form $1/2+m$, with $m=0$;  thus in order to fit the
first case $\hat{\lambda}$ must be of the form
$1/2+l$, $l$ an integer, that is
$$
\frac{1}{2}\sqrt{1+4 \omega ^2}=\frac{1}{2}+\ell
$$
 this yields the condition
\begin{equation}\label{appendix-(ii)}
\omega^2=\ell(\ell+1)
\end{equation}
Now observe that condition (\ref{appendix-(i)}) is contained in condition (\ref{appendix-(ii)})
by taking $l=2n-1$ or $l=2n$.

\qed
\begin{remark}
We recovered condition (\ref{condn}).
\end{remark}

\section*{Acknowledgments}
Primitivo Acosta--Hum\'anez is partially supported by grant FPI
Spanish Government, project BFM2003-09504-C02-02 and is grateful
to the Department of Mathematics of  Universidad Aut\'onoma
Metropolitana--Iztapalapa for the hospitality during the stay of
research for the development of this work. Martha
\'Alvarez--Ram\'\i rez and Joaqu\'in Delgado were partially
supported by CONACYT-M\'exico, grant 47768 and by a PIFI 2007
project UAM-I-CA-55 "Differential Equations and Geometry"; the
last author thanks the hospitality of Universidad Sergio Arboleda
in Bogot\'a, where this research was continued, and specially to
Reinaldo N\'u\~nez head of the Mathematics Department for his kind
support.


\newpage
\begin{thebibliography}{9}
\bibitem{AbrahamMarsden}
R.~Abraham and J.~E. Marsden, \emph{Foundations of mechanics},
  Benjamin/Cummings Publishing Co. Inc. Advanced Book Program, Reading, Mass.,
  1978, Second edition, revised and enlarged, With the assistance of Tudor Ra\c
  tiu and Richard Cushman.

\bibitem{Abramowitz}
M.~Abramowitz and I.~A. Stegun (eds.), \emph{Handbook of
mathematical functions
  with formulas, graphs, and mathematical tables}, A Wiley-Interscience
  Publication, John Wiley \& Sons Inc., New York, 1984, Reprint of the 1972
  edition, Selected Government Publications.

\bibitem{ac}
P.~B. Acosta-Humanez, \emph{Non-Autonomous Hamiltonian
Systems and Morales-Ramis Theory I. The Case \lowercase{$\ddot
x=f(x,t)$}}, to appear in SIAM Journal on Applied Dynamical
Systems.

\bibitem{acbl}
P.~B. Acosta-Humanez and D. Bl\'azquez-Sanz,
\emph{Non-integrability of some hamiltonians with rational
potentials},  Discrete Contin. Dyn. Syst. Ser. B  \textbf{10}
(2008), no. 2-3, 265--293.

\bibitem{Arnold}
V.~I. Arnold, \emph{Mathematical methods of classical
mechanics},Springer-Verlag, New York, 1978, Translated from the Russian by K. Vogtmann and A. Weinstein, Graduate Texts in Mathematics, 60.

\bibitem{Boucher} D. Boucher, \emph{Sur la non-int\'egrabilit\'e du probl\`eme
plan des trois corps de masses \'egales}, C. R. Acad. Sci. Paris S\'er. I Math. \textbf{331} (2000), no.~5, 391--394.

\bibitem{BoucherWeil}
D. Boucher and J.-A. Weil, \emph{Application of J.-J. Morales and
J.-P. Ramis'theorem to test the non-complete integrability of the
planar three-body problem}, From combinatorics to
  dynamical systems, IRMA Lect. Math. Theor. Phys., vol.~3, de Gruyter, Berlin, 2003, pp.~163--177.

\bibitem{dulo} A. Duval and M. Loday-Richaud, \emph{Kovacic's algorithm and its application to some families of special functions},  Appl. Algebra Engrg. Comm. Comput. \textbf{3} (1992), no. 3, 211--246.

\bibitem{Kov} J. Kovacic, \emph{An Algorithm for Solving Second Order Linear Homogeneus Differential Equations, }
J. Symb. Comput. {\bf 2} (1986), 3--43.

\bibitem{kimura} T. Kimura, \emph{On Riemann's Equations which are Solvable by Quadratures},
Funkcialaj Ekvacioj {\bf 12} (1969), 269-281.

\bibitem{Morales}
J.~J. Morales-Ruiz, \emph{Differential {G}alois theory and
non-integrability
  of {H}amiltonian systems}, Progress in Mathematics, vol. 179, Birkh\"auser
  Verlag, Basel, 1999.

\bibitem{mo1} J.~J. Morales-Ruiz,
\emph{A remark about the {P}ainlev\'e transcendents}, Th\'eories
  asymptotiques et \'equations de Painlev\'e, S\'emin. Congr. Angiers, vol.~14,
  Soc. Math. France, Paris, 2006, pp.~229--235.

\bibitem{MoralesSplitting}
J.~J. Morales-Ruiz and J.~M. Peris, \emph{On a Galoisian approach
to the splitting of separatrices}, Ann. Fac. Sci. Toulouse Math.
(6) \textbf{8}
  (1999), no.~1, 125--141.

\bibitem{moralesramis}
J.~J. Morales-Ruiz and J.~P. Ramis, \emph{Galoisian obstructions
to integrability of Hamiltonian systems. I}, Methods Appl. Anal.
\textbf{8}  (2001), no.~1, 33--96.

\bibitem{moralesramisII}
J.~J. Morales-Ruiz and J. P. Ramis, \emph{Galoisian obstructions
to integrability of Hamiltonian systems. II}, Methods Appl. Anal.
\textbf{8} (2001), no.~1, 97--112.

\bibitem{MorRamSimo}
J. J. Morales Ruiz, J. P. Ramis and C. Sim\'o, \emph{Integrability
of Hamiltonian Systems and Differential Galois Groups of Higher
Variational Equations}, Ann. Sci. École Norm. Sup. (4) \textbf{40}
(2007), no. 6, 845--884.

\bibitem{simon}
S. Simon i Estrada, \emph{On the Non-integrability of some
Problems in Celestial Mechanics}, Vdm Verlag Dr Mueller EK, Berlin, 2008.

\bibitem{SingerVanderput}
M. van~der Put and M.~F. Singer, \emph{Galois theory of
linear differential equations}, Grundlehren der Mathematischen Wissenschaften, vol. 328, Springer-Verlag,
Berlin, 2003.

\bibitem{UlmerWeil} F. Ulmer and J.-A. Weil, \emph{Note on Kovacic's algorithm}.
J. Symb. Comp. \textbf{22}, (1996), 179--200.

\bibitem{szebe} V. G. Szebehely, \emph{Theory of Orbits}, Academic Press. New York and London, 1966.

\bibitem{vigo}
M. I. Vigo-Aguiar, M.E. Sansaturio, J.M. Ferr\'andiz,
\emph{Integrability of Hamiltonians with polynomial potentials},
Journal of Computational and Applied Mathematics {\bf 158},
(2003), 213-224.

\bibitem{Whittaker}
E.~T. Whittaker, \emph{A treatise on the analytical dynamics of
particles and rigid bodies}, Cambridge Mathematical Library, Cambridge University Press,
Cambridge, 1988, With an introduction to the problem of three bodies, Reprint
of the 1937 edition, With a foreword by William McCrea.

\bibitem{yagasaki}
K. Yagasaki, \emph{Galoisian obstructions to integrability
and Melnikov criteria for chaos in two-degree-of-freedom
Hamiltonian systems with saddle centres}, Nonlinearity \textbf{16}
(2003), no. 6, 2003--2012.

\bibitem{yoshida} H. Yoshida, {\em A criterion for the non-existence of an additional integral in
hamiltonian systems with a homogeneous potential}, Physica D {\bf 29} (1987),
128-142.

\bibitem{ziglin} S.L. Ziglin, {\em Bifurcation of solutions and the nonexistence of first integrals in Hamiltonian mechanics. I. (Russian)}
 Funktsional. Anal. i Prilozhen. \textbf{16} (1982), no. 3,
 30--41.

\end{thebibliography}
\end{document}